%% file: main.tex
\documentclass[conference]{IEEEtran}

\usepackage{blindtext, graphicx,  mathtools, amsmath}
\usepackage{cite}

\usepackage{scrextend}
\ifCLASSINFOpdf
\else
\fi

\begin{document}
%
\title{Blink Rate Variability during resting and reading sessions}

\input{sec_authors.tex}

\maketitle

\section*{Abstract}
\input{sec_abstract.tex}

%
\IEEEpeerreviewmaketitle

\section{Introduction}
\input{sec_introduction.tex}

\section{Materials and methods}
\input{sec_materials.tex}

\section{Procedure}
\input{sec_estimation.tex}


\section{Results and discussion}
\input{sec_results.tex}

\section{CONCLUSION}
\input{sec_conclusion.tex}

\setcounter{secnumdepth}{0}
\section{Appendix}
\input{sec_appendix1.tex}

\ifCLASSOPTIONcaptionsoff
  \newpage
\fi

\input{sec_references.tex}
\end{document}

%% file: sec_authors.tex


%
\author{\IEEEauthorblockN{
Artem Lenskiy, Rafal Paprocki}

\IEEEauthorblockA{Korea University of Technology and Education\\
1600, Chungjeol-ro, Byeongcheon-myeon,
Dongnam-gu, Cheonan-si, Chungcheongnam-do 31253,\\ Republic of Korea\\
email: lensky@koreatech.ac.kr, rafal.paprocki@gmail.com}
}



%% file: sec_abstract.tex
\begin{abstract}
It has been shown that blinks occur not only to moisturize eyes and as a defensive response to the environment, but are also caused by mental processes. In this paper, we investigate statistical characteristics of blinks and blink rate variability of 11 subjects. The subjects are presented with a reading/memorization session preceded and followed by a resting session. EEG signals were recorded during these sessions. The signals from the two front electrodes were then analyzed, and times of the blinks were detected. We discovered that compared to the resting sessions, reading session is characterized by a lower number of blinks. However, there was no significant difference in standard deviation in the blink rate variability. We also noticed that in terms of complexity measures, the blink rate variability is located somewhere in between white and pink noises, being closer to the white noise during reading. We also found that the average of inter-blink intervals increases during reading/memorization, thus longer inter-blink intervals could be associated with a mental workload.

\end{abstract}
\begin{IEEEkeywords}
eye blink, Blink Rate Variability, mental stages, rest and read
\end{IEEEkeywords}

%% file: sec_introduction.tex
Majority of foregoing research has focused on eye blinks occurrence during performing some particular tasks or being exposed to external stimuli. Researchers suggested that blinking can be an indicator of the transition of one cognitive process to another, in other words, that eye blinks can be markers of the beginning and ending of a certain cognitive process. In \cite{10} it has been shown that blink rate patterns are influenced more by cognitive processes rather than by age, eye color or gender.\\
Interior brain activities have significant influence on the blinks. Ponder and Kennedy concluded, that high-level cognitive processes are major determinants of a blink increase and reduction \cite{4}. They also highlighted that blinks could be an indicator of attention, named by them as “mental tension.”  Operational memory, which is used during performing mental tasks and visual imagination, may share components with a visual perceptual system. To avoid interference of cognitive processes, blinking is slowed down \cite{16}.\\
It has been reported that blinks play an important role in detecting various brain disorders and are helpful in distinguishing brain activities. Spontaneous Blink Rate (BR) has been studied in context of many neurological diseases like Parkinson's disease, schizophrenia or Tourette syndrome \cite{5}\cite{6}\cite{7}\cite{2}. It has been discovered that in detecting psychiatric disorders like schizophrenia and attention hyperactivity BR can be helpful as a source of data. It is possible because blinks are identified as non-invasive peripheral markers of the central dopamine activity. That makes their accurate detection and analysis useful.\\ 
Research on blink duration and reopening time during the morning and evening sessions shows that during evening, durations of both characteristics are significantly longer \cite{21}.\\
Studies suggested that blink behaviour during reading is under perceptual and cognitive control \cite{17}. Furthermore, in 1972 Holland and Tarlow \cite{15} not only defined blinking as a function of memory load, "such that the more the items in memory (and presumably the more the rehearsal activity), the fewer the blinks," but also suggested that blinking can be an interlude between ideas or sentences \cite{16}. \\
Researchers have shown the synchronous behaviour in blinking between a listener and a speaker in face-to-face conversation. 
Workload in some visuospatial tasks can be indicated from multiple eye measures, such as blink frequency and duration of eye fixation at the point of interest, during a task in which memory and visual activity demands vary over time \cite{20}. BR is inversely correlated to the increase of workload, so blinks can be also used to detect drowsiness \cite{14}. \\
Every time brain activates for cognitive processes, resources are consumed \cite{26}. Resource is identified as the level of activation in the underlying cortical neural system, available for processing the information and storing it \cite{27}. Resource management is considered as a source of individual differences in performance of the cognitive process. 
For the resource allocation measurements, various methods can be used\cite{26}, among which eye blink analysis still remains understudied. 
Difficulties in an interpretation of eye blinks as a reflection of mental state reveal that blink can appear in natural response of eye dryness, reflex, or other environmental stimuli. 
Therefore, eye inter-blink dynamics could be a better indicator of relationship of eyelid activity and mental states. Among a number of characteristics of dynamical systems or measurements of dynamical system outputs, analysis of correlation properties plays an important role.   \\ 
One type of correlation is a long-range correlation, long-range correlation effects have been found in various processes including biological, physiological, and financial. These processes, having self-affinity (self-similarity, SS) properties, and are characterized by the scale exponents. 
Some specific examples of processes with LRC include the Internet traffic and particularly WiMAX network traffic \cite{ALenskiy2}. The scale exponents are good characteristics to discern cases of heart failure and healthy heart behaviour \cite{Peng2}. The scale exponents of heartbeat dynamics and respiration dynamics are different for young and elderly people \cite{ALenskiy1}.\\
It is evident that eye blinks are linked to brain activity. In this paper, we investigate the relationship between the blink rate variability and two mental activities: reading and resting. We state that the same tasks involve alike parts of brain. Therefore, characteristics of blinks have to be the same in respect of subjects. \\
The experimental setup is described in section 2. To estimate blink rate variability, we utilize a blink detection algorithm that we proposed earlier \cite{refHK} and its description is given in section 3, together with Hurst exponent's calculation method. Section 4 discusses the relationship between the blink rate variability and the task (reading and resting). Finally, section 5 concludes our study.    

%% file: sec_materials.tex
\subsection{Data acquisition}
While participants were performing the test we recorded the videos with a Pointgrey Flea3 high frame rate USB camera. We stored the video for future work. Simultaneously, EEG signals were captured using Mitsar-EEG 201 amplifier and accompanying WinEEG software. The electrodes were placed according to the international “10-20 system” standards of electrode placement. 
Electro-gel was injected into electrodes hollow in order to decrease the electrode-skin resistance. The focus of current work is to analyze the statistical characteristics of blinks. Nevertheless, we record EEG from all the electrodes for further research. 

\begin{figure}[!htbp]
\includegraphics[width=3.5in]{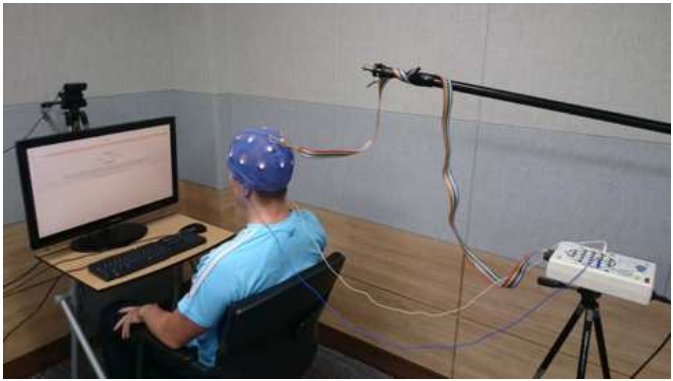}
\caption{Setup of the experiment}
\label{fig2}
\end{figure}

\subsection{Testing procedure}

\begin{figure}[!htbp]
\includegraphics[trim=0.0cm 1.5cm 0.0cm 2.5cm, width=3.5in]{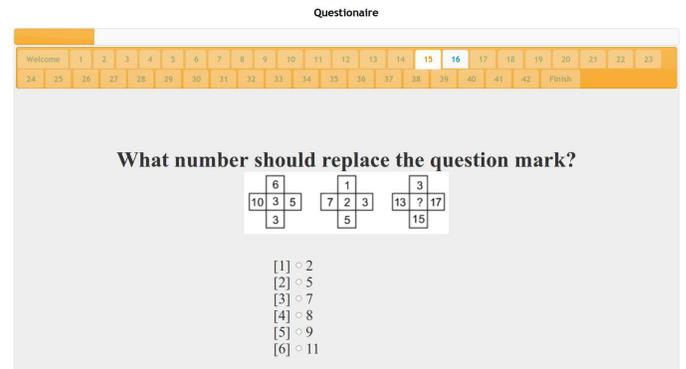}
\caption{Question delivery software}
\label{fig5a}
\end{figure}

Fifty one male and female subjects aged from 19 to 25 years, were recruited for the experiment. Everyone provided their written consent. The subjects had no history of psychiatric illness, and they had not been affected by any significant medical, neurological or ophthalmological illness. We prepared a questionnaire testifying that subjects had none of these conditions. 
The experiment software was developed in JavaScript with jQuery (Fig. \ref{fig5a}). The software was designed in a way that  no intervention of  participants or an experiment supervisor is required during the procedure. \\
The whole testing session took 30 minutes and consisted of 5 minute resting, followed by a 10 minute IQ testing, and a 5 minute rest. After the second resting period, a passage about Ethiopia was presented, followed by a quiz to invoke  memory recall activity. The first 5 minute resting period was included to calm the participants down. Participants were asked to try to keep their mind calm as much as possible. The following IQ test is not discussed in this paper as more investigation is needed. The latter 5 minute resting stage is included in this study and the comparison between the two resting sessions is provided in later sections. The passage is presented after the second resting interval for five minutes. Participants were asked to read the passage carefully with the purpose of memorizing the facts from it. The passage is 40 sentences long, and contains basic facts about Ethiopia. None of the participants were familiar with the topic of the text. The passage followed by the 5 minute quiz to test the memory retention, we  will discuss it in our next study. \\
Since EEG is prone to noise due to movement, we made sure that subjects had been informed about staying still and their movement was minimized. During both resting stages, information with instruction about staying still with countdown progress bar, were displayed. To prevent head movements, text was displayed on a single screen with no need to scroll. Due to heavy noise caused by the subjects falling asleep, adjusting the cap or constant head movements, the total of 40 subjects' data were dropped.\\ 

%% file: sec_estimation.tex
To begin analysis the statistical characteristics of blinks, we have to precisely detected positions of the blinks in the recorded EEGs. Among all channels, we only  analyzed signals recorded from the electrodes placed near the eyes, as our focus is detection of the electrical potential generated by muscle responsible for blinks. Then the recordings were imported as CSV files to Matlab for further analysis. The process of blink detection has been previously described \cite{refHK}. Here we briefly outline it.
We divided the blink detection procedure into two parts: the preprocessing part and the blink detection part. The preprocessing consists of the following four steps: (a) bandpass filtering to remove harmonics that are not caused by blinks, (b) cut off extremely high and deeply low amplitudes, such amplitudes were estimated from the cumulative distribution function, (c) apply independent component analysis and remove the component with brain activity. The blink detection part is performed on the preprocessed signals and consists of (d) signal thresholding, (e) blink candidate extraction, (f) polynomial fitting and maximum localization (fig. 4). After detecting the blink, blink rate variability (BRV) is calculated as differences in time of consecutive blinks. (fig. 5).
The process of BRV extraction is followed by a statistical analysis. The main goal of the analysis is to examine the statistical characteristics of blink-rate variability and check, whether depending on the type of a mental activity, statistical properties of BRV change. Among the statistics we have investigated, the type of correlation BRV is one of the important ones. We estimated scale exponent for the BRV, as it is known to characterize complexity and correlation of random processes. To estimate scale exponent, we applied an estimation algorithm that improves on the well-known Detrended Fluctuation Analysis method \cite{DFA1}. The DFA suffers from the overestimation problem \cite{ALenskiy3} that is a result of the polynomial detrending.  For the purpose of suppressing unwanted harmonics outside of the analyzed frequency band, instead of subtracting a polynomial that is a nonlinear operator\cite{DFA1} we applied band-pass filtering that is a linear operation \cite{ALenskiy1}. In order to estimate scale exponents a range of scales have been selected. The frequency range is proportional to the analyzed scale $l$. We set the range of scales as  $l = \{4,	5,	6,	7,	8,	10,	11,	13,	16,	19,	23,	27, 32\}$. The maximum scale is chosen in a  way to be smaller than at least half of the shortest BRV's length among all participants and across all stages. We estimated the scale exponent $\alpha$ for each of the subjects for the resting and reading stages.
\begin{figure}[!htbp]
\includegraphics[trim=0.0cm 0.5cm 0.0cm 0.5cm, width=3.5in]{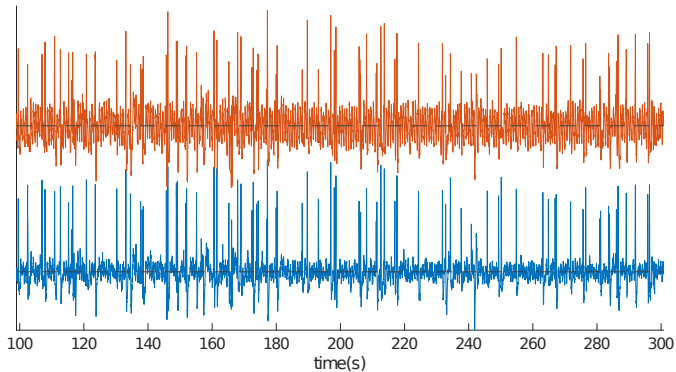}
\caption{ Fp1-Fp3 and Fp2-Fp4 electrode pairs}
\label{fig3}
\end{figure}

\begin{figure}[!htbp]
\includegraphics[trim=0.0cm 0.5cm 0.0cm 0.5cm, width=3.5in]{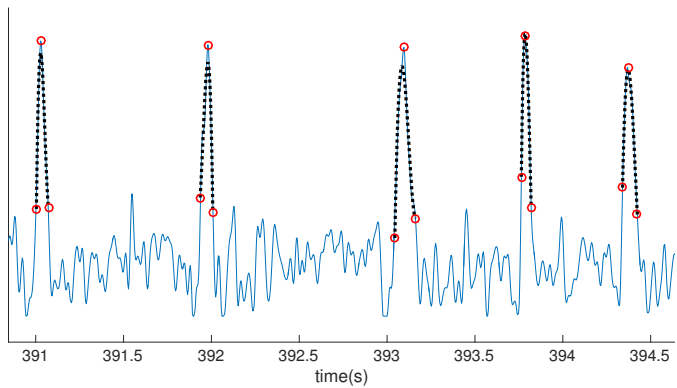}
\caption{3rd degree polynomial function is fitted to every waveform blink candidate. The localized peak corresponds to the time of a blink.}
\label{fig4}
\end{figure}

\begin{figure}[!htbp]
\includegraphics[trim=0.0cm 0.5cm 0.0cm 0.0cm, width=3.5in]{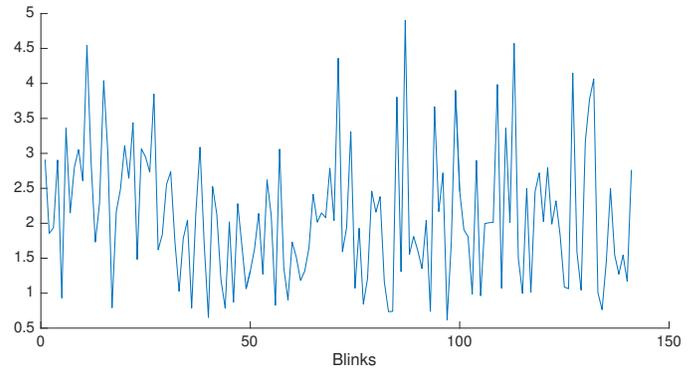}
\caption{Extracted blink rate variability}
\label{fig5}
\end{figure}

%% file: sec_results.tex
In 1927 Ponder and Kennedy \cite{4} reported that higher mental processes were major determinants of blink enhancements and inhibition. 
They also mentioned that blinks might serve as an index of attention, or as they termed it “mental tension”. They inferred that person inhibits blinking while actively being engaged into information abstraction. Interesting results were presented by Hall \cite{19}, who concluded that blinks do not occur randomly during reading. 
Bentivoglio et al, studied BR during rest, reading and conversation, the authors investigated the blink rate (BR) patterns; they showed that the patterns are influenced more by cognitive processes rather than by age, eye color or gender \cite{10}. 

\begin{figure}[!htbp]
\includegraphics[trim=0.0cm 0.25cm 0.0cm 0.5cm, width=3.5in]{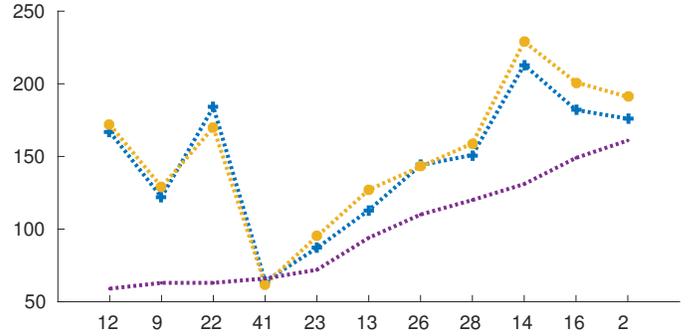}
\caption{Number of blinks depending on the task - blue and yellow plots are resting stages 1 and 2 respectively, purple plot is reading stage}
\label{fig6}
\end{figure}

\begin{figure}[!htbp]
\includegraphics[width=3.5in]{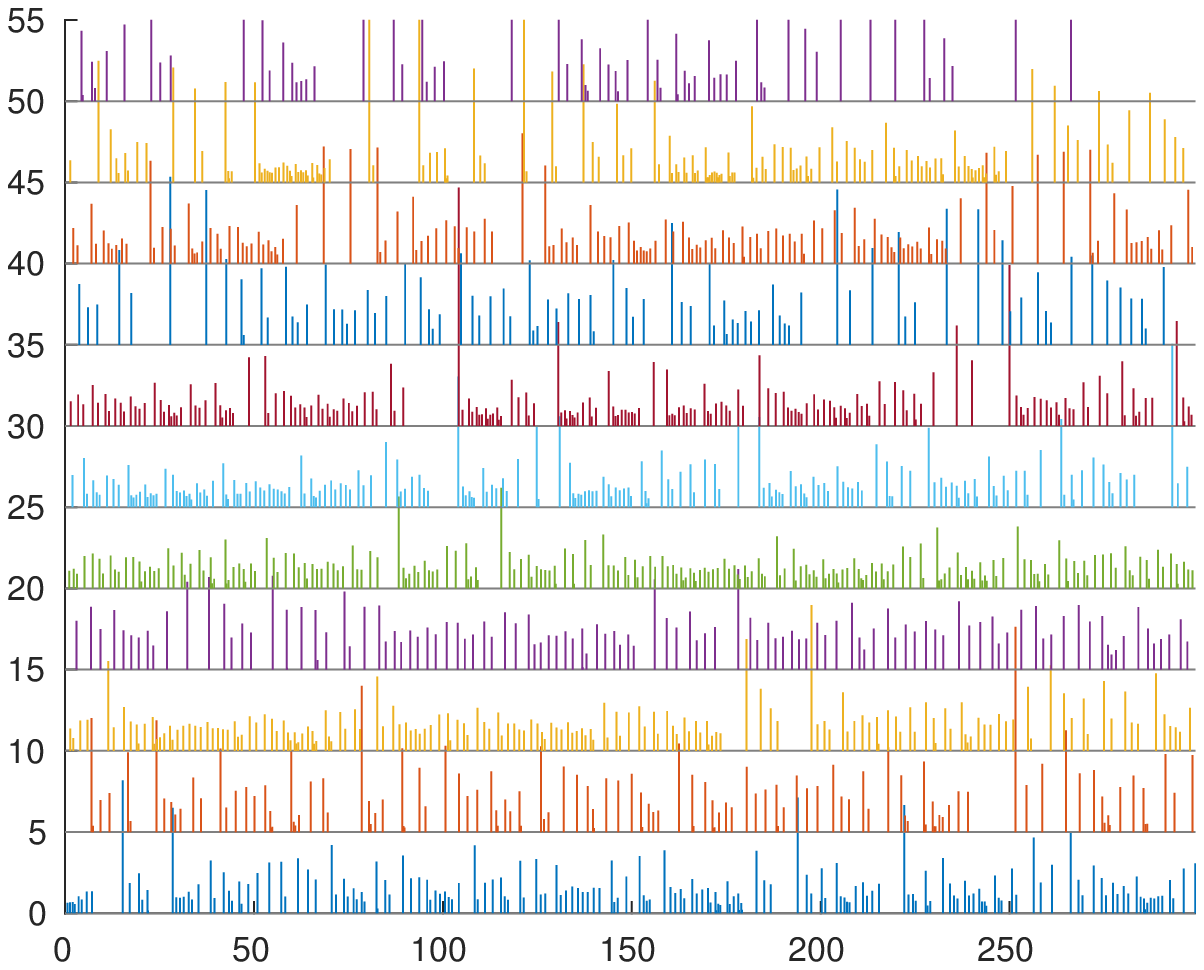}
\caption{BRV during first rest stage for all subjects; y axis is interval length, x axis is blink time}
\label{fig7}
\end{figure}

\begin{figure}[!htbp]
\includegraphics[width=3.5in]{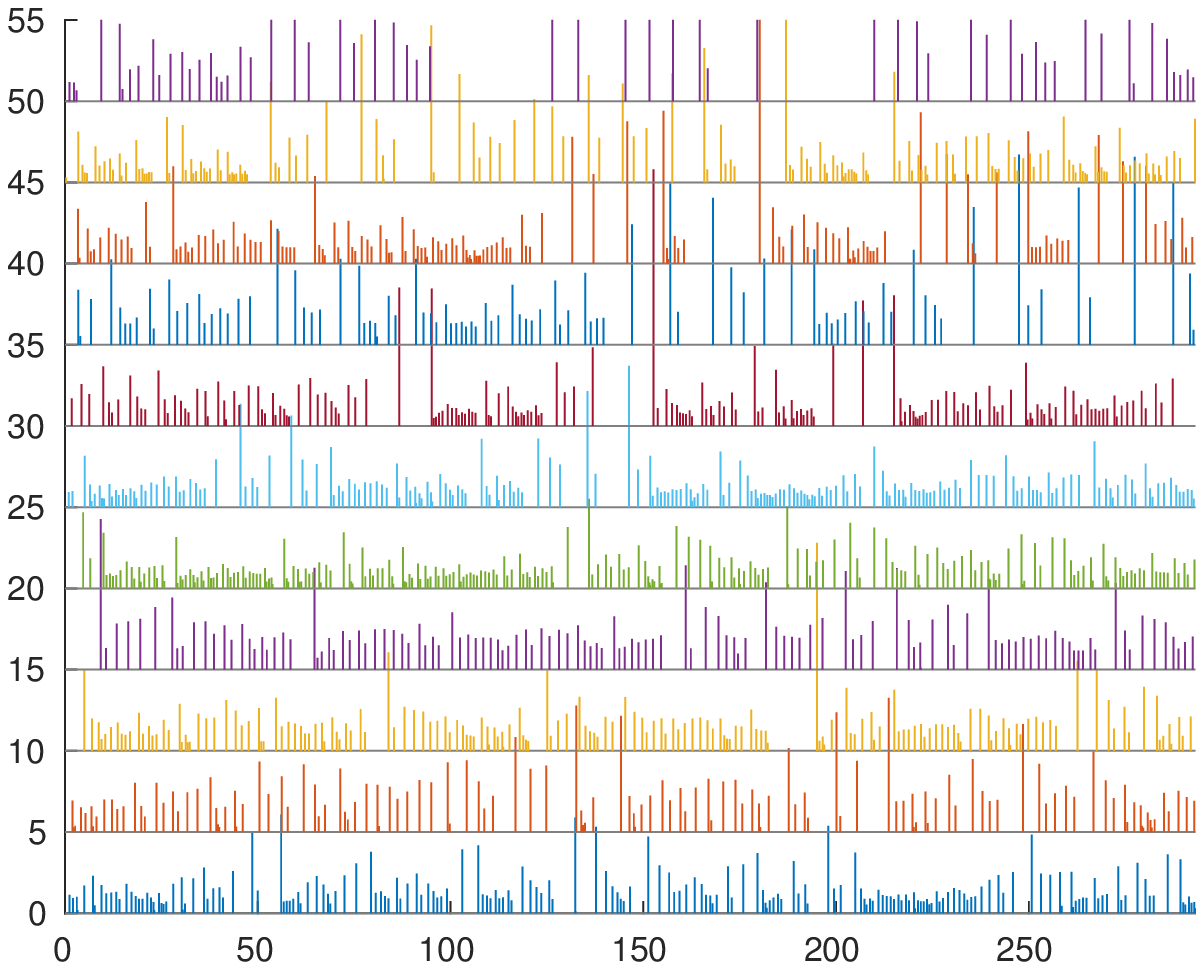}
\caption{BRV during second rest stage for all subjects; y axis is interval length, x axis is blink time}
\label{fig8}
\end{figure}

\begin{figure}[!htbp]
\includegraphics[trim=0.0cm 0.25cm 0.0cm 0.5cm, width=3.5in]{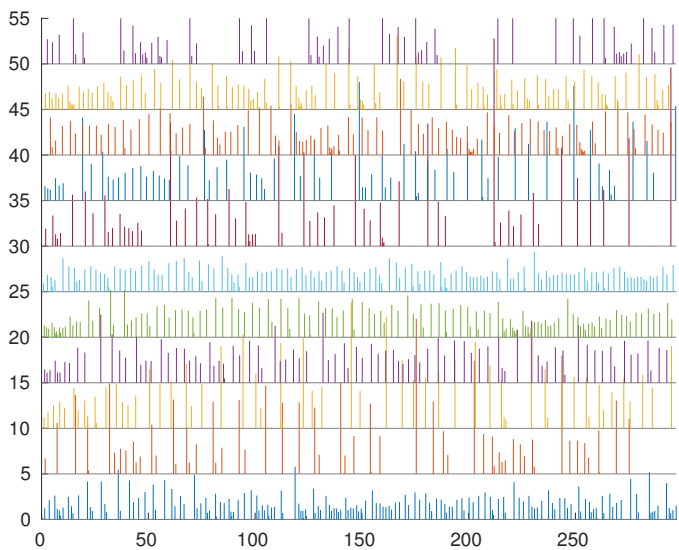}
\caption{BRV during reading stage for all subjects; y axis is interval length, x axis is blink time}
\label{fig9}
\end{figure}

\begin{figure}[!htbp]
\includegraphics[trim=0.0cm 0.25cm 0.0cm 0.5cm, width=3.5in]{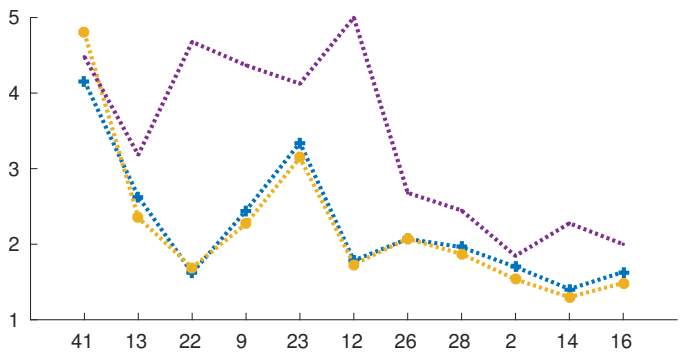} 
\caption{Average inter-blink - blue and yellow plots are resting stages 1 and 2 respectively, purple plot is reading stage}
\label{fig10}
\end{figure}

\begin{figure}[!htbp]
\includegraphics[trim=0.0cm 0.25cm 0.0cm 0.0cm, width=3.5in]{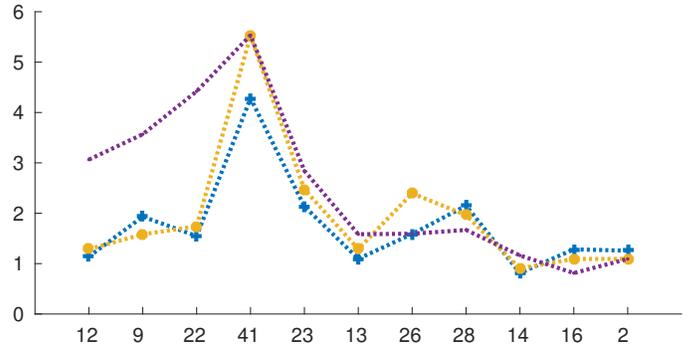}
\caption{Standard deviation of IBI dynamics - blue and yellow plots are resting stages 1 and 2 respectively, purple plot is reading stage}
\label{fig11}
\end{figure}

\begin{figure}[!htbp]
\includegraphics[trim=0.0cm 0.0cm 0.0cm 0.0cm, width=3.5in]{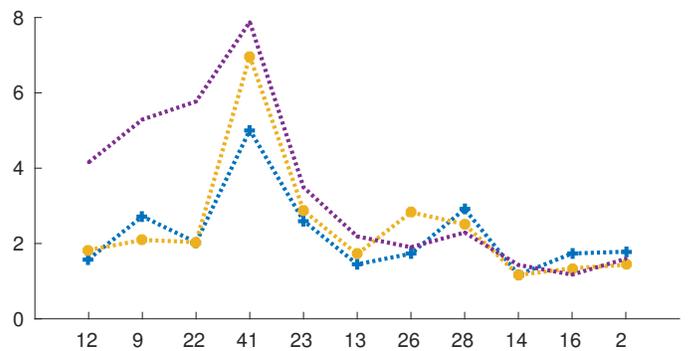}
\caption{Root Mean Square of the Successive Differences of IBI - blue and yellow plots are resting stages 1 and 2 respectively, purple plot is reading stage}
\label{fig12}
\end{figure}

Holland et. al reported that blinking is related to certain cognitive processes \cite{15,16}. The authors stated that blinking rate decreases as mental load increases. Subjects were asked to perform simple arithmetic calculations on a sequence of numbers. As the sequence elongated, so presumably the memory load increased, number of blinks decreased. In our experiment, reading the text plays a role of a workload. The number of blinks per each of 11 subjects is presented in Fig \ref{fig6}. Comparing the blinking rates during both resting sessions reveals they are significantly closer to each other rather than during resting and reading sessions. It can be seen that blinking is indeed an indicator of a mental state, the number of blinks during reading is smaller than during resting sessions for 10 out of 11 subjects. That consistency of blinking ratio during different tasks dictates the existence of blinking patterns. Indeed, in \cite{10} authors have shown, that for 92\% of the population, a higher blinking rate was observed during resting than reading.\\
On the other hand, time between blinks does change. The average inter blink interval is shorter for resting and elongates for reading. One of the explanations of longer inter-blink intervals in reading is fixation points that occur while tracking the text. It is necessary to understand visual perception and eye movement in order to understand the reading process. Reading involves a prolonged focus on reading material as each word is related to its predecessors to make sense of the whole sentence. Furthermore, reading is a process in which the eyes quickly move to assimilate the text. From one fixation to another, there is a search for a next fixation thus this requires less blinking and more focus.
For visually demanding tasks, blinks rate is lower and positioned in time to maximize stimulus detection (Fournier 1999, Goldstein 1992, Ponder 1928).
Reading is a cognitive process that increases brain activity \cite{17} which leads to longer inter-blink intervals. Another possible explanation for longer intervals is a reflection of the blinking concept as an interlude between ideas or sentences \cite{16}.\\
Figures \ref{fig7}, \ref{fig8} and \ref{fig9} show blink rate variability for all subjects during resting sessions and reading, respectively. The abscissa is the blink interval number whereas the ordinate represents the interval lengths per each subject. 
The mean of BRV is always higher during reading, except of one subject (fig. \ref{fig10}). Fig. \ref{fig11}, presents the standard deviation (SD) of the BRV dynamics. The root mean square of the successive differences (RMSSD) of the BRV is shown in fig. \ref{fig12} and calculated as:

\begin{equation}
 RMSSD = \sqrt{\frac{1}{N-1} \sum_{n=1}^{N-1}(I_n - I_{n-1})^2)},
\label{eq:1}
\end{equation}
where $I_n$ is length of the $n$th interblink interval. 

The RMSSD is often employed in analysing the HRV, where every sample is an interval between successive heartbeats. The RMSSD characterizes HR's short-term variations, and its value is low during subject's high stress \cite{31}. In the case of blinks, the number of blinks decreases during reading, whereas average inter-blink intervals increases. The RMSSD is noticeably higher during reading session for 6 out of 11 subjects, for the remaining subjects, there are no noticeable differences. That could suggest either unusual behaviour of fixation phenomena, or different resource management strategy that depends on an individual.\\
As for the next step, we evaluate the complexity of the BRV and investigate if it is affected by various types of mental workload \cite{29}.
Many physiological processes are statistically self-similar(SS). One of the definitions of a SS process $X(t)$ is given via its power spectral density and defined as

\begin{equation}
    S(f) = \frac{\sigma_{X}}{ f^{\gamma}}
\label{eq:2}
\end{equation}

\begin{figure}[!htbp]
\includegraphics[trim=0.0cm 0.25cm 0.0cm 0.0cm, width=3.5in]{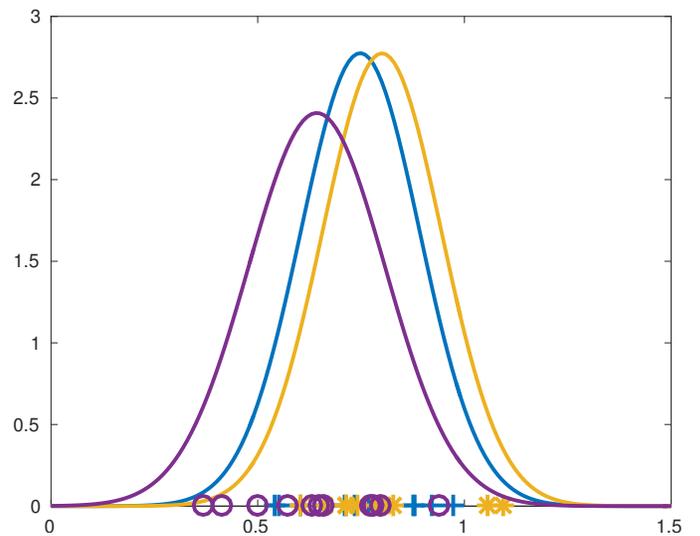}
\caption{Gaussian PDFs with the mean and standard deviations estimated for the Hurst exponents of IBI during resting sessions 1 (blue) and 2 (yellow) and for the reading session (purple).}
\label{fig13}
\end{figure}

The complexity of statistically self-similar processes is directly determined by the spectral exponent $\gamma$. Scale exponents are the parameters that characterise the rate of scaling. Notice that according to the definition \ref{eq:2}, the following property holds $S(a \cdot f)=a^{-\gamma} \cdot \sigma_{X}/ f^{\gamma}$.  One of the methods to estimate scale exponent is the detrended fluctuation analysis (DFA). Unfortunately, the method is known to suffer from overestimation problems\cite{ALenskiy2}. To improve the estimation accuracy, we employ a modified DFA \cite{ALenskiy3}. The method estimates the scale exponent $alpha$ that is related to the spectral exponent $\gamma$ via $\alpha = (\gamma + 1)/2$. Taking the advantage of this method, we estimate scale exponents for the BRV of every subject per every session. The mean $\langle \alpha \rangle$ and the standard deviation $ \sigma $ of the estimated scale exponents are shown in table \ref{table1}. In order to interpret the complexity of the BRV, we compare it to color noise which power spectral density is given by \ref{eq:2}. Following this definition, white noise corresponds to the process with  $\alpha = 0.5$ i.e. a random process with a constant spectral density. Pink noise is characterized by the spectral density $S(f) \propto 1/f$ i.e $ \alpha = 1$. It is known that many physical and physiological processes are of $1/f$ type, some examples include body movements, walking, HRV, brain dynamics and cognition \cite{34}. Brown noise corresponds to $\alpha = 1.5$. It is widely accepted that pink noise is of highest complexity, whereas the complexity of brown noise decays as it is a cumulative sum of discrete white noise. It is worth noticing that spectral characteristics of the HRV for young people with a healthy cardio-vascular system correspond to pink noise, whereas for elderly people the dynamics of HRV is shifting towards brown noise. Scale exponents estimated for the BRV are distributed between white and pink noises.  We plot Gaussian probability density functions with estimated means and standard deviations for each of these sessions and visualized them in fig. \ref{fig13}. It is easy to see, that the means of both of the resting stages ($0.75\pm0.14$ and $0.80\pm0.14$) are located closer to each ($\delta = 0.05$), while the difference between the  means for the resting sessions to the reading session is located farther apart ($\delta = 0.10$ and $\delta = 0.15$ correspondingly). Thus, we can conclude, the complexity of the BRV varies among different tasks. Furthermore, during the resting sessions, the complexity of BRV increases as it shifts towards pink noise. \\


\begin{table}[]
\centering
\caption{Hurst exponent}
\label{table1}
\begin{tabular}{llllll}
\hline

\begin{tabular}[c]{@{}l@{}}Subject/session \end{tabular} & 
\begin{tabular}[c]{@{}l@{}}Rest session 1\end{tabular} & 
\begin{tabular}[c]{@{}l@{}}Resting session 2\end{tabular} & 
\begin{tabular}[c]{@{}l@{}}Reading session\end{tabular} & 
\\ \hline
1 &	    0.73 &	    0.8 &	0.41  \\
2 &	    0.55 &	    0.71 &	0.57  \\
3 &	    0.97 &	    0.8 &	0.66  \\
4 &	    0.54 &	    0.6 &	0.63  \\
5 &	    0.66 &	    0.66 &	0.78  \\
6 &	    0.6 &	    1.09 &	0.37  \\
7 &	    0.92 &	    1.05 &	0.78  \\
8 &	    0.71 &	    0.74 &	0.5  \\
9 &	    0.88 &	    0.79 &	0.94  \\
10 &	    0.88 &	    0.71 &	0.8  \\
11 &	    0.77 &	    0.83 &	0.65  \\
\hline
$\langle \alpha \rangle \pm \sigma $
& $0.75 \pm 0.14 $
& $0.80 \pm 0.14 $
& $0.65 \pm 0.17 $
\\

\hline
\end{tabular}
\end{table}

%% file: sec_conclusion.tex
We have shown that the act of reading is associated with a smaller number of blinks compared to the resting sessions. Reading has lower blink rate because people read sentences by fixating on words. Our findings support \cite{17} that blinking behavior during reading is under perceptual and cognitive control. With different experimental setup,
our finding is concurrent with the findings of \cite{18} that reports on a significant and a positive correlation between inter-blink intervals and subsequent memory encoding. We observed a higher blink rate while subjects were resting. Furthermore, the distribution of $\alpha$'s which characterize complexity is varies among sessions depending on a task. This suggests that type of mental activity affects blink-rate variability and its complexity. 
In our future work, we are planning to extend our work to more types of mental activities and investigate on the relationship between statistical characteristics of the BRV and mental activities  

%% file: sec_appendix1.tex
All used abbreviations are expanded below:

\begin{labeling}{abbreviations}
\item [BR] Blink Rate
\item [BRV] Blink Rate Variability
\item [IBI] Inter-blink Intervals
\item [DFA] Detrended Fluctuation Analysis 
\item [EEG] Electroencephalogram
\item [HR] Heart Rate
\item [HRV] Heart Rate Variability
\item [RMSSD] Root  Mean  Square  of  the Successive  Differences
\item [SD] Standard Deviation
\item [SS] Self Similarity

\end{labeling}